\documentclass{aipproc}

\newcommand{\be}{\begin{equation}}
\newcommand{\ee}{\end{equation}}
\newcommand{\bea}{\begin{eqnarray}}
\newcommand{\eea}{\end{eqnarray}} \newcommand{\nn}{\nonumber}

\newcommand{\de}{\partial}

\layoutstyle{6x9}

\SetInternalRegister\hbadness{8000} 

\newcommand\doingARLO[2][]{%
  \ifx\mmref\undefined #1\else #2\fi
}

\begin{document}
\def\esp #1{e^{\displaystyle{#1}}}
\def\slash#1{\setbox0=\hbox{$#1$}#1\hskip-\wd0\dimen0=5pt\advance
       \dimen0 by-\ht0\advance\dimen0 by\dp0\lower0.5\dimen0\hbox
         to\wd0{\hss\sl/\/\hss}}\def\ink {\int~{d^4k\over (2\pi)^4}~}

\title
      [Effective lagrangians for QCD at high density]
      {Effective lagrangians for QCD at high density}


\author{Roberto Casalbuoni}{
  address={Dipartimento di Fisica dell' Universita' di Firenze and Sezione
INFN, L.go E. Fermi 2, 50125 Firenze, Italy. E-mail: casalbuoni@fi.infn.it},
  email={casalbuoni@fi.infn.it},
}

\begin{abstract}
We describe low energy physics in the CFL and
LOFF phases by means of effective lagrangians. In the CFL case we
present also how to derive expressions for the parameters
appearing in the lagrangian via weak coupling calculations
taking advantage of the dimensional reduction of fermion physics
around the Fermi surface. The Goldstone boson of the LOFF phase
turns out to be a phonon satisfying an anisotropic dispersion
relation.
\end{abstract}

\date{\today}

\maketitle

\section{Introduction} Ideas about color superconductivity
go back to almost 25 years ago \cite{barrois}, but only recently
this phenomenon has received a lot of attention (for recent
reviews see ref. \cite{wilczek1}). The naive expectation is that
at very high density, due to the asymptotic freedom, quarks would
form a Fermi sphere of almost free fermions. However, Bardeen,
Cooper and Schrieffer proved that the Fermi surface of free
fermions is unstable in presence of an attractive, arbitrary
small, interaction.  Since in QCD  the gluon exchange in the
$\bar 3$ channel is attractive one expects the formation of a
coherent state of particle/hole  pairs (Cooper pairs). For a
careful description of the formation of the condensates and of the
approximations involved in going from asymptotic densities to
finite ones it is useful to see the contribution of K. Rajagopal
at this meeting \cite{rajagopal1}. The phase structure of QCD at
high density depends on the number of flavors and there are two
very interesting cases, corresponding to two massless flavors
(2SC) \cite{barrois,2SC} and to three massless flavors (CFL)
\cite{CFL1,CFL2} respectively. In this talk we will be mainly
concerned with the latter case. The two cases correspond to very
different patterns of symmetry breaking. If we denote left- and
right-handed quark fields by $\psi_{iL(R)}^\alpha$ with
$\alpha=1,2,3$, the $SU(3)_c$ color index, and $i=1,\cdots,N_f$
the flavor index ($N_f$ is the number of massless flavors), in the
2SC phase we have the following structure for the condensate
($C=i\gamma^2\gamma^0$ is the charge-conjugation matrix) \be
\langle q_{iL(R)}^\alpha C q_{jL(R)}^\beta\rangle\propto
\epsilon_{ij} \epsilon^{\alpha\beta3}. \label{2SC} \ee The
condensate breaks the  color group $SU(3)_c$ down to the subgroup
$SU(2)_c$ but it does not break any flavor symmetry. Although the
baryon number, $B$, is broken, there is a combination of $B$ and
of the broken color generator, $T_8$, which is unbroken in the 2SC
phase. Therefore no massless Goldstone bosons are present in this
phase. On the other hand, five gluon fields acquire mass whereas
three are left massless. It is worth to notice that  for the
electric charge the situation is very similar to the one for the
baryon number. Again a linear combination of the broken electric
charge and of the broken generator $T_8$ is unbroken in the 2SC
phase. The condensate (\ref{2SC})  gives rise to a gap, $\Delta$, for quarks
of color 1 and 2, whereas the two quarks of color 3 remain
un-gapped (massless). The resulting effective low-energy theory
has been described in \cite{sannino}. In this contribution we
will be mainly interested in the formulation of the effective
theory for the three massless quarks case. At high density it has
been shown that the following condensate is formed
\cite{CFL1,CFL2} \be \langle q_{iL(R)}^\alpha C
q_{jL(R)}^\beta\rangle\propto \epsilon^{ijX} \epsilon_{\alpha\beta
X}+\kappa(\delta^i_\alpha\delta^j_\beta+\delta_\beta^i\delta_\alpha^j)
\label{CFL}. \ee Due to the Fermi statistics, the condensate must
be symmetric in color and flavor. As a consequence the two terms
appearing in eq. (\ref{CFL}) correspond to the $(\mathbf{\bar
3},\mathbf{\bar 3})$ and $(\mathbf{6},\mathbf{6})$ channels of
$SU(3)_c\otimes SU(3)_{L(R)}$. It turns out that $\kappa$ is small
\cite{CFL1, small1,small2} and therefore the condensation occurs
mainly in the $(\mathbf{\bar 3},\mathbf{\bar 3})$ channel. The
expression (\ref{CFL}) shows that the ground state is left
invariant by a simultaneous transformation of $SU(3)_c$ and
$SU(3)_{L(R)}$. This is called Color Flavor Locking (CFL). The
symmetry breaking pattern is
\bea &SU(3)_c\otimes SU(3)_L\otimes SU(3)_R\otimes U(1)_B\otimes
U(1)_A&\nn\\ &\downarrow&\\ &SU(3)_{c+L+R}\otimes Z_2\otimes
Z_2&\nn \eea The $U(1)_A$ symmetry is broken at the quantum level
by the anomaly, but it gets restored at very high density since
the instanton contribution is suppressed \cite{anomaly,
small1,son1}. The $Z_2$ symmetries arise since the condensate is
left invariant by a change of sign of the left- and/or
right-handed fields. As for the 2SC case the electric charge is
broken but a linear combination with the broken color generator
$T_8$ annihilates the ground state. On the contrary the baryon
number is broken. Therefore there are $8+2$ broken global
symmetries giving rise to 10 Goldstone bosons. The one associated
to $U(1)_A$ gets massless only at very high density. The color
group is completely broken and all the gauge particles acquire
mass. Also all the fermions are gapped. We will show in the following how to
construct an effective
lagrangian describing the Goldstone bosons, and how to compute
their couplings in the high density limit where the QCD coupling
gets weaker. A final problem we will discuss has to do
with the fact that when quarks (in particular the strange quark)
are massive, their chemical potentials cannot be all equal. This
situation has been modeled out in \cite{LOFF1}.  If the Fermi
surfaces of different flavors are too far apart, BCS pairing does not occur.
However it
might be favorable for different quarks to pair each of one lying
at its own Fermi surface and originating a pair of non-zero  total
momentum. This is the LOFF  state first studied by the authors of
ref. \cite{LOFF2} in the context of electron superconductivity in
the presence of magnetic impurities. Since the Cooper pair has
non-zero momentum the condensate breaks space symmetries and
we will show that  in the low-energy spectrum a massless
particle, a phonon, the Goldstone boson of the broken
translational symmetry, is present. We will construct the effective lagrangian
also for this case.

\section{Effective theory for the CFL phase}

We start introducing the Goldstone fields as the phases of the
condensates in the $(\mathbf{\bar 3},\mathbf{\bar 3})$ channel
\cite{casa1,hong1}\be X_\alpha^i\approx
\epsilon^{ijk}\epsilon_{\alpha\beta\gamma}\langle q^j_{\beta
L}q^k_{\gamma L}\rangle^*,~~~ Y_\alpha^i\approx
\epsilon^{ijk}\epsilon_{\alpha\beta\gamma}\langle q^j_{\beta
R}q^k_{\gamma R}\rangle^*. \ee Since quarks belong to the
representation $(\mathbf{3},\mathbf{3})$ of $SU(3)_c\otimes
SU(3)_{L(R)}$ and transform under $U(1)_B\otimes U(1)_A$
according to \be q_L\to e^{i(\alpha+\beta)} q_L,~~~q_R\to
e^{i(\alpha-\beta)} q_R,~~~e^{i\alpha}\in U(1)_B,~~~e^{i\beta}\in
U(1)_A, \ee the transformation properties of the fields $X$ and
$Y$ under the total symmetry group $G=SU(3)_c\otimes
SU(3)_L\otimes SU(3)_R\otimes U(1)_B\otimes U(1)_A$ are ($g_c\in
SU(3)_c$, $g_{L(R)}\in SU(3)_{L(R)}$) \be X\to g_cXg_L^T
e^{-2i(\alpha+\beta)},~~~Y\to g_cYg_R^T e^{-2i(\alpha-\beta)}. \ee
The fields $X$ and $Y$ are $U(3)$ matrices and as such they
describe $9+9=18$ fields. Eight of these fields are eaten up by
the gauge bosons, producing eight massive gauge particles.
Therefore we get the right number of Goldstone bosons, $10
=18-10$. These fields correspond to the breaking of the global
symmetries in $G$ (18 generators) to the symmetry group of the
ground state $H=SU(3)_{c+L+R}\otimes Z_2\otimes Z_2$ (8
generators). For the following it is convenient to separate the
$U(1)$ factors in $X$ and $Y$ defining fields belonging to
$SU(3)$ \be X=\hat X e^{2i(\phi+\theta)},~~~Y=\hat Y
e^{2i(\phi-\theta)},~~~\hat X,\hat Y\in SU(3). \ee The fields
$\phi$ and $\theta$ can  also be described  through the
determinants of $X$ and $Y$ \be d_X={\rm
det}(X)=e^{6i(\phi+\theta)},~~~d_Y={\rm
det}(Y)=e^{6i(\phi-\theta)}, \ee The transformation properties
under $G$ are\be \hat X\to g_c\hat X g_L^T,~~~\hat Y\to g_c\hat Y
g_R^T,~~~\phi\to\phi-\alpha,~~~\theta\to\theta-\beta. \ee The
breaking of the global symmetry can   be discussed in terms of gauge
invariant fields given by $d_X$, $d_Y$
and\be\Sigma^i_j=\sum_\alpha (\hat Y_\alpha^j)^* \hat
X_\alpha^i\to \Sigma=\hat Y^\dagger \hat X. \ee The $\Sigma$
field describes the 8 Goldstone bosons corresponding to the
breaking of the chiral symmetry $SU(3)_L\otimes SU(3)_R$, as it
is made clear by the transformation properties of $\Sigma^T$,
$\Sigma^T\to g_L \Sigma^T g_R^\dagger$. That is $\Sigma^T$
transforms exactly as the usual chiral field. The other two fields
$d_X$ and $d_Y$ provide the remaining two Goldstone bosons related
to the breaking of the $U(1)$ factors.

In order to build up an invariant lagrangian, it is convenient to
define the following currents \be J_X^\mu=\hat X D^\mu \hat
X^\dagger=\hat X(\partial^\mu\hat X^\dagger+\hat X^\dagger
g^\mu),~~~J_Y^\mu=\hat Y D^\mu \hat Y^\dagger=\hat
Y(\partial^\mu\hat Y^\dagger+\hat Y^\dagger g^\mu), \ee with
$g_\mu=ig_s g_\mu^a T^a/2$ the gluon field and $T^a=\lambda_a/2$
the $SU(3)_c$ generators. These currents have simple
transformation properties under the full symmetry group $G$,
$J^\mu_{X,Y}\to g_c J^\mu_{X,Y}g_c^\dagger$. The most general
lagrangian, up to two derivative terms, invariant under $G$, the
rotation group $O(3)$ (Lorentz invariance is broken by the
chemical potential term) and the parity transformation defined as
$\hat X\leftrightarrow \hat Y$, $\phi\to\phi$, $\theta\to
-\theta$, is \cite{casa1} \bea {\cal L}&=&-\frac{F_T^2}4{\rm
Tr}\left[\left(J^0_X-J^0_Y)^2\right)\right]-\alpha_T\frac{F_T^2}4{\rm
Tr}\left[\left(J^0_X+J^0_Y)^2\right)\right]+\frac 12
(\de_0\phi)^2+\frac
12(\de_0\theta)^2\nn\\&&\label{lagrangian1}\\
&&+\frac{F_S^2}4{\rm Tr}\left[\left(\vec J_X-\vec
J_Y)^2\right)\right]+\alpha_S\frac{F_S^2}4{\rm
Tr}\left[\left(\vec J_X+\vec
J_Y)^2\right)\right]-\frac{v_\phi^2}2|\vec\nabla\phi|^2-
\frac{v_\theta^2}2|\vec\nabla\theta|^2.\nn
 \eea Using $SU(3)_c$ color gauge invariance
we can choose $\hat X=\hat Y^\dagger$, making 8 of the Goldstone
bosons  disappear and giving mass to the gluons. The properly
normalized Goldstone bosons, $\Pi^a$,  are given in this gauge by
\be \hat X=\hat Y^\dagger =e^{i\Pi^a T^a/F_T}, \ee and expanding
eq. (\ref{lagrangian1}) at the lowest order in the fields we get
\be {\cal L}\approx\frac 12 (\de_0\Pi^a)^2+\frac 12
(\de_0\phi)^2+\frac 12(\de_0\theta)^2 -\frac{v^2}
2|\vec\nabla\Pi^a|^2-\frac{v_\phi^2}2|\vec\nabla\phi|^2-
\frac{v_\theta^2}2|\vec\nabla\theta|^2,\ee with $v=F_s/F_T$. The
gluons $g_0^a$ and $g_i^a$ acquire Debye and Meissner masses
given by \be m_D^2=\alpha_Tg_s^2 F_T^2,~~~m_M^2=\alpha_Sv^2g_s^2
F_T^2. \label{masses}\ee It should be stressed that these are not
the true rest masses of the gluons, since there is a large wave
function renormalization effect making the gluon masses of order
of the gap $\Delta$, rather than $\mu$ (see later) \cite{casa2}.
Since this description is supposed to be valid at low energies (we
expect much below the gap $\Delta$), we could also decouple the
gluons solving their classical equations of motion neglecting the
kinetic term. The result from eq. (\ref{lagrangian1}) is \be
g_\mu=-\frac 12 \left(\hat X\de_\mu\hat X^\dagger+\hat
Y\de_\mu\hat Y^\dagger\right). \ee It is easy to show that
substituting this expression in eq. (\ref{lagrangian1}) one gets
\cite{casa2} \be {\cal L}=\frac{F_T^2}4\left({\rm
Tr}[\dot\Sigma\dot\Sigma^\dagger]-v^2{\rm
Tr}[\vec\nabla\Sigma\cdot\vec\nabla\Sigma^\dagger]\right)+ \frac
12\left(\dot\phi^2-v_\phi^2|\vec\nabla\phi|^2\right)+ \frac
12\left(\dot\theta^2-v_\phi^2|\vec\nabla\theta|^2\right). \ee
Notice that the first term is nothing but the chiral lagrangian
except for the breaking of the Lorentz invariance. This is a way
of seeing the quark-hadron continuity, that is the continuity
between the CFL and the nuclear matter phases in three flavor QCD.
The identification is perfect if one realizes that in nuclear
matter the pairing may occur in such a way to give rise to a
superfluid due to the breaking of the baryon number as it happens
in the CFL phase \cite{schafer2}.

\section{Fermions near  the Fermi surface}

We will introduce now the formalism described in ref.
\cite{hong2} in order to evaluate the parameters  appearing in the
effective lagrangian. This formulation is based on the
observation that, at very high-density, the energy spectrum of a
massless fermion is described by states $|\pm\rangle$ with
energies  $E_\pm =-\mu\pm|\vec p$ where $\mu$ is the quark number
chemical potential. For energies much lower than the Fermi energy
$\mu$, only the states $|+\rangle$ close to the Fermi surface.
i.e. with  $|\vec p\,|\approx\mu$, can be excited. On the
contrary, the states $|-\rangle$ have $E_-\approx -2\mu$ and
therefore  decouple. This can be seen more formally by writing the
four-momentum of the fermion as \be p^\mu=\mu
v^\mu+\ell^\mu\label{momentum},\ee where $v^\mu=(0,\vec v_F)$,
and $\vec v_F$ is the Fermi velocity defined as $\vec v_F=\de
E/\de\vec p|_{\vec p=\vec p_F}$. For massless fermions $|\vec
v_F|=1$. Since the hamiltonian for a massless Dirac fermion in a
chemical potential $\mu$ is \be H=-\mu+\vec\alpha\cdot\vec
p\,,~~~\vec\alpha=\gamma_0\vec\gamma, \ee one has \be
H=-\mu(1-\vec\alpha\cdot \vec v_F)+\vec \alpha\cdot\vec\ell.\ee
Then, it is convenient to introduce the projection operators \be
P_\pm=\frac{1\pm\vec\alpha\cdot\vec v_F}2,\ee such that \be
H|+\rangle=\vec\alpha\cdot\vec\ell\, |+\rangle,~~~
H|-\rangle=(-2\mu+\vec\alpha\cdot\vec\ell )\,|-\rangle.\ee

We can define
fields corresponding to the states $|\pm\rangle$ through the
decomposition \be \psi(x)=\sum_{\vec v_F}\,\esp{-i\mu v\cdot
x}\left[\psi_+(x)+\psi_-(x)\right],\ee where  an average over the
Fermi velocity $\vec v_F$ is
performed. The velocity-dependent fields $\psi_\pm (x)$ are given
by ($v^\mu=(0,\vec v_F)$)
 \be \psi_\pm(x)=\esp{i\mu v\cdot
x}\left(\frac{1\pm\vec\alpha\cdot\vec
v_F}2\right)\psi(x)~=~\int_{|\ell
|<\mu}\frac{d^4\ell}{(2\pi)^4}\esp{-i\,\ell\cdot
x}\psi_\pm(\ell).\label{psimeno}\ee Since we are interested at
physics near the Fermi surface we integrate out all the modes
with $|\ell|>\mu$. Substituting inside the Dirac part of the QCD
lagrangian density one obtains ($V^\mu=(1,\vec v_f)$, $\tilde
V^\mu=(1,-\vec v_F)$)\be {\cal L}=\sum_{\vec v_F}
\left[\psi_+^\dagger iV\cdot D\psi_++\psi_-^\dagger(2\mu+ i\tilde
V\cdot D)\psi_-+(\bar\psi_+i\slash D_\perp\psi_- + {\rm
h.c.})\right],\ee where $\slash D_\perp=D_\mu\gamma^\mu_\perp$ and
\be
\gamma^\mu_\perp~=~\frac 1 2 \gamma_\nu\left(2 g^{\mu\nu}-
V^\mu\tilde V^\nu-\tilde V^\mu V^\nu \right).
\ee

We notice that the fields appearing in this
expression are evaluated at the same Fermi velocity because
off-diagonal terms are canceled by the rapid oscillations of the
exponential factor in the $\mu\to\infty$ limit. This behavior
can be referred to as the Fermi velocity super-selection rule.

At the leading order in $1/\mu$ one has
 \be iV\cdot
D\psi_+=0,~~~~\psi_-= -\frac{i}{2\mu}\gamma_0\slash D_\perp\psi_+,\ee showing
the decoupling of $\psi_-$ for $\mu\to\infty$. The
equation for $\psi_+$ shows also that only the energy and the
momentum parallel to the Fermi velocity are relevant variables in
the problem. We have an effective two-dimensional theory.

At the next to leading order the effective action for the field $\psi_+$ is
\be
{\cal L}=\sum_{\vec v_F}\left[\psi_+^\dagger iV\cdot D\psi_+-\frac
1{2\mu}\psi_+^\dagger(\slash D_\perp)^2\psi_+\right].\label{effective}
\ee

The previous remarks apply to any theory describing massless
fermions at high density. The next step will be to couple this
theory in a $SU(3)_L\otimes SU(3)_R\otimes SU(3)_c$ invariant way
to  Nambu-Goldstone bosons (NGB)  describing the appropriate
breaking  for the CFL phase (we will not discuss here the determination of the
parameters relevant for the $U(1)_{A,B}$ fields, see \cite{son1}). Using a
gradient expansion we
get an explicit  expression for  the decay coupling constant of
the Nambu-Goldstone boson as well for their velocity.

The invariant coupling between fermions and Goldstone fields
reproducing the symmetry breaking pattern of eq. (\ref{CFL}) is
proportional to \be \gamma_1\,Tr[\psi_L^T\hat X^\dagger]\,C\,
Tr[\psi_L \hat X^\dagger]+\gamma_2\,Tr[\psi_L^TC\hat
X^\dagger\psi_L \hat X^\dagger]+{\rm h.c.},\label{invariant}\ee
with an analogous expression for the right-handed fields. Here the
spinors are meant to be Dirac spinors. The trace is operating
over the group indices of the spinors and of the Goldstone
fields. Since the vacuum expectation value of the Goldstone
fields is $\langle \hat X\rangle=\langle \hat Y\rangle=1$, we see
that this coupling induces the right breaking of the symmetry. In
the following we will consider only the case
$\gamma_2=-\gamma_1\propto\Delta/2$, where $\Delta$ is the gap
parameter.

Since the transformation properties under the symmetry group of
the fields at fixed Fermi velocity do not differ from those of the  quark
fields, for both
left-handed and right-handed fields we get the effective
lagrangian density \bea {\cal L}&=&\sum_{\vec v_F} \frac 1 2
\Big[\sum_{A=1}^9\left(\psi_+^{A\dagger}iV\cdot
D\psi_+^A+\psi_-^{A\dagger}i\tilde V\cdot
D\psi_-^A-{\Delta_A}\left({\psi_-^A}^TC\psi_+^A+{\rm
h.c.}\right)\right)\cr&-&{\Delta}\sum_{I=1,3}\left(Tr[(\psi_-
X_1^\dagger)^T C\epsilon_I(\psi_+
X_1^\dagger)\epsilon_I]+\rm{h.c.}\right)\Big],
\label{interaction}\eea where we have introduced the fields $\psi_\pm^A$:
\be\psi_\pm=\frac{1}{\sqrt{2}}\sum_{A=1}^9\lambda_A\psi^A_\pm.\ee
Here $\lambda_a~(a=1,...,8)$ are the Gell-Mann matrices normalized
as follows: $Tr(\lambda_a\lambda_b)=2\delta_{ab}$ and
$\lambda_9=\sqrt{ 2/ 3}~{\bf {1}}$. Furthermore
$\Delta_1=\cdots=\Delta_8=\Delta$, $\Delta_9=-2\Delta$, and
$X_1=\hat X-1$. Notice that  the NGB fields couple to fermionic fields
with opposite Fermi velocities. In this expression, as in the
following ones, the field $\psi_-$ is defined as $\psi_+$ with
$\vec v_F\to-\vec v_F$, and therefore it is not the same as the
one defined in (\ref{psimeno}).

The formalism becomes more compact by introducing the
Nambu-Gorkov fields \be \chi=\left(\matrix{\psi_+\cr
C\psi^*_-}\right).\ee It is important to realize that the fields
$\chi$ and $\chi^\dagger$ are not independent variables. In fact,
since we integrate over all the Fermi surface, the fields
$\psi_-^*$ and $\psi_+$, appearing in $\chi$, appear also in
$\chi^\dagger$ with $\vec v_F\to -\vec v_F$. In order to avoid
this problem we can integrate over half of the Fermi surface, or,
taking into account the invariance under $\vec v_F\to -\vec v_F$,
we can simply integrate over all the sphere with a weight $1/8\pi$
instead of $1/4\pi$. Then the first three terms in the lagrangian
density (\ref{interaction}) become \be {\cal L}_0=\int\frac{d\vec
v_F}{8\pi}~ \frac 1 2\sum_{A=1}^9
\chi^{A\dagger}\left[\matrix{iV\cdot D & \Delta^A\cr\Delta^A
&i\tilde V\cdot D^*}\right]\chi^A,\label{kinetic}\ee so that, in
momentum space the free  fermion propagator is \be
S_{AB}(p)=\frac{2\delta_{AB}}{V\cdot p\,\tilde V\cdot
p-\Delta_A^2}\left[\matrix{\tilde V\cdot p & -\Delta_A\cr-\Delta_A
& V\cdot p}\right].\ee

 We are now in position to evaluate the
self-energy of the Goldstone bosons through their coupling to the
fermions at the Fermi surface. There are two one-loop
contributions \cite{casa2}, one from the coupling $\Pi\chi\chi$
and a tadpole from the coupling $\Pi\Pi\chi\chi$ (see eq.
(\ref{interaction}). The tadpole diagram contributes only to the
mass term and it is essential to cancel the external momentum
independent  term arising from the other diagram. Therefore, as
expected, the mass of the NGB's is zero. The contribution at the
second order in the momentum expansion is given by
 \be
i\,  \frac{21-8\ln 2}{72\pi^2 F_T^2} \int\frac{d\vec
v_F}{4\pi}\sum_{a=1}^8 \Pi^a\,V\cdot p\, \tilde V\cdot p\,\Pi^a.
\ee Integrating over the velocities and going back to the
coordinate space we get \be{\cal L}_{\rm eff}^{\rm kin}=
\frac{21-8\ln 2} {72\pi^2 F_T^2}\sum_{a=1}^8
\left(\dot\Pi^a\dot\Pi^a-\frac 1 3|\vec\nabla\Pi_a|^2\right). \ee
We can now determine the decay coupling constant $F_T$ through
the requirement of getting the canonical normalization for the
kinetic term; this implies \be F_T^2=\frac {\mu^2(21-8\ln
2)}{36\pi^2},\ee a result  obtained by many authors using
different methods (for a complete list of the relevant papers see
the first reference of \cite{wilczek1}). We see also that
$v^2=1/3$. The other constants appearing in our effective
lagrangian can be obtained via a direct calculation of $m_D$ and
$m_M$ \cite{casa2}. This is done evaluating the one-loop
contribution to the gluon self-energy. Also in this case there
are two contributions, one coming from the gauge coupling to the
fermions, whereas the other arises from the next-to leading (in
$\mu$) sea-gull contribution to the fermion effective lagrangian
in eq. (\ref{effective}). The results we find are \cite{casa2,son1} \be
m_D^2=g_s^2 F_T^2,~~~ m_M^2=\frac 1 3 m_D^2. \ee Comparison with
equation (\ref{masses}) shows that \be \alpha_S=\alpha_T=1. \ee
Performing a gradient expansion of the gluon self-energy one
finds that there is a wave function renormalization of order
$g_s\mu /\Delta \gg 1$. Extrapolating this result,  the physical
masses of the gluons turn out to be of the order of the gap
energy ($\approx 1.70\Delta$) \cite{casa2}.
 The origin of the pion velocity
$1/\sqrt{3}$ is a direct consequence of the integration over the
Fermi velocity. Therefore it is completely general and applies to
all the NGB's in the theory, including the ones associated to the
breaking of $U(1)_V$ and $U(1)_A$ ($v_\phi^2=v_\theta^2=1/3$); needless to say
that higher
order terms in the expansion $1/\mu$ could change this result.

The breaking of the Lorentz invariance exhibited by the pion
velocity different from one, can be seen also in the matrix
element $\langle 0|J_\mu^a|\Pi^b\rangle$. Its evaluation gives \cite{casa2}
\be\langle 0|J_\mu^a|\Pi^b\rangle=iF_T\delta_{ab} \tilde p_\mu,
~~~~\tilde p^\mu=(p^0, \vec p/3).\ee The current is conserved, as a
consequence of the dispersion relation satisfied by NGB's.

\section{The LOFF phase}
We shall now consider massless quarks of  three colors and two
different flavors. At finite densities we introduce two chemical
potentials, $\mu_1$ and $\mu_2$, for the two species in order to
mimic the different mass case. We write
\be \mu_1-\mu_2=\delta
\mu\ll\mu=\frac{\mu_1+\mu_2}2. \ee
The BCS condensation takes place also  for $\delta\mu\neq 0$
provided $\delta\mu\ll \Delta $. On the other hand, for
$\delta\mu\approx\Delta $, the picture changes significantly. The
analysis in \cite{LOFF1} shows that there exist two values of
$\delta\mu$, $\delta\mu_1$ and $\delta\mu_2$, such that, for $
\delta\mu\in(\delta\mu_1,\delta\mu_2)$ the high density
quark-gluon matter is in a phase characterized by the breaking of
translational and rotational invariance, due to the presence of a
scalar and a vector condensate. This phenomenon is called
crystalline color superconductivity of QCD and the relative
phase is named LOFF phase.  The authors of ref. \cite{LOFF1} find
$\delta\mu_1=0.71\Delta$ and $\delta\mu_2=0.744\Delta$ for
$\mu=0.4~GeV$ and $\Delta=40~MeV$ for a point-like four-fermi
coupling. More recently it has been found that in the
one-gluon exchange approximation the window opens up considerably
\cite{LOFF3}.

The condensation in the LOFF phase gives rise to two breaking terms in the
fermion lagrangian characterized by two gap parameters $\Delta^{(s)}$ and
$\Delta^{(v)}$
 \be-\frac 1 2\, e^{2i\vec q\cdot\vec x}\sum_{\vec
v_F}e^{i\delta\mu\vec v_F\cdot\vec x}\left[\Delta^{(s)}\epsilon_{ij}+\vec
v\cdot\vec
n\Delta^{(v)}\sigma^1_{ij}\right]\epsilon^{\alpha\beta
3}\psi_{+\vec v;\,i\alpha}\,C\,\psi_{-\vec v;\,j\beta}-(L\to R),
\label{ldeltaeff}\ee with $\vec n=\vec q/|\vec q|$. The
condensates break the space symmetry group. However
the discussion of the number of NGB's in the case of space
symmetries is a subtle one due to the particular group structure.
In fact rotations and translations cannot be considered
transformations breaking the symmetries of the theory in an
independent way. This is because a translation plus a rotation is
physically equivalent to a translation.  Let us discuss the
consequences of this situation more closely.

 We first consider spatial rotations.
 We define a
vector field $\vec R(x)$ such that $|\vec R|^2=1$ and $\langle
\vec R\rangle_0=\vec n$ \cite{casa3}. The rotational symmetry is
restored by substituting $\vec v\cdot\vec n\to \vec v\cdot\vec R$
in the term proportional to $\Delta^{(v)}$.  Let us now consider
the exponential factor
 $\exp(2 i\vec q \cdot\vec x)$
 in (\ref{ldeltaeff}), which
  breaks both rotational and translational invariance.
By introducing a field $\Phi(x)$ behaving as a scalar under the
space group \cite{casa3}, we restore translational and rotational
invariance via  the substitution $ 2\vec q\cdot\vec x\to\Phi(x)$.
We assume $\langle\Phi(x)\rangle_0=2\vec q\cdot\vec x$. We then
introduce a field $\phi(x)$ through \be\ \Phi(x)=2\vec q\cdot\vec
x+\phi(x),~~~\langle\phi(x)\rangle_0=0, \ee and convenient
transformation properties such to compensate the variation of the
term $\vec q\cdot\vec x$ under the space group \cite{casa3}. The
field $\phi$ acts as the phonon (Nambu-Goldstone boson) field
associated to the breaking of the space symmetry. We can now
construct the field $\vec R$ in terms of $\Phi(x)$ as $\vec R
=\vec \nabla\Phi/|\vec \nabla\Phi|$. This expression  satisfies
the required properties for $\vec R(x)$.

Through a bosonization procedure similar to the one employed in
the previous Section, one can derive an effective lagrangian for
the NGB field. The effective lagrangian must contain only
derivative terms. Polynomial terms are indeed forbidden by
translation invariance, since $\phi$ is not a scalar field under
space transformations. In order to write the kinetic terms is
better to use the  field $\Phi$ which behaves as a scalar under
both rotations and translation.  However since the expectation
value of the gradient of $\Phi$ is given by
$\langle\vec\nabla\Phi\rangle_0=2\vec q\approx \Delta$, we cannot limit the
expansion in the spatial derivatives of $\Phi$ to any finite
order. A real spatial derivative expansion can be made only for
the phonon field $\phi$. With this in mind the most general
invariant lagrangian will contain a tower of space-derivative
terms \cite{casa3}:
 \be{\cal
L}(\phi,\partial_\mu\phi)=\frac{f^2}
2\left[\dot\Phi^2-\sum_{n=1}^\infty c_n
(|\vec\nabla\Phi|^{2})^n\right].\label{Phi}\ee Here $\Phi$ must be
thought as a function of the phonon field $\phi$. Using \be
|\vec\nabla\Phi(x)|^2=4q^2+\frac{4q} f\vec
n\cdot\vec\nabla\phi(x)+\frac 1{f^2}|\vec\nabla\phi(x)|^2,\ee at
the lowest order in the derivatives of the phonon field $\phi$ we
get (neglecting a constant term): \be{\cal
L}(\phi,\partial_\mu\phi)=\frac{1}{2}\left[\dot\phi^2-
v_\parallel^2|\vec\nabla_\parallel\phi|^2-
v^2\left(4qf\vec\nabla_\parallel\phi+|\vec\nabla\phi|^2\right)
\right],\label{phonon}\ee where $\vec\nabla_\parallel\phi=\vec
n\cdot\vec\nabla\phi$ and $v_\parallel^2$,  $v^2$ are constants.
Notice that the linear term gives rise to a surface contribution.
The lack of rotational invariance in (\ref{phonon}) follows from
the gradient expansion due to the non-linear transformations
undergone by the field $\phi(x)$. This happens also in the
analogous expansion for the chiral field. Therefore the physical
consequence of the extraction of the expectation value of $\Phi$
is  an anisotropy in the dispersion relation for the phonon field
$\phi(x)$.

\begin{theacknowledgments}
I would like to thank R. Gatto, M. Mannarelli  and  G. Nardulli for their
precious collaboration to the papers originating this contribution.

\end{theacknowledgments}


\end{document}